\documentstyle[12pt]{article}
\newcommand{\be}{\begin{equation}}
\newcommand{\ee}{\end{equation}}
\newcommand{\bea}{\begin{eqnarray}}
\newcommand{\eea}{\end{eqnarray}}

\newcommand{\p}[1]{(\ref{#1})}
\newcommand{\lb}{\label}

\begin{document}
\begin{titlepage}
\begin{flushright}
January 2008
\end{flushright}
\vskip 0.6truecm

\begin{center}
{\Large\bf Gauging isometries in ${\cal N}=4$ supersymmetric mechanics}
\vspace{1.5cm}

{\large\bf F. Delduc$\,{}^a$, E. Ivanov$\,{}^b$,}\\
\vspace{1cm}

{\it a)ENS Lyon, Laboratoire de Physique, 46, all\'ee d'Italie,}\\
{\it 69364 LYON Cedex 07, France}\\
{\tt francois.delduc@ens-lyon.fr}
\vspace{0.3cm}

{\it b)Bogoliubov  Laboratory of Theoretical Physics, JINR,}\\
{\it 141980 Dubna, Moscow region, Russia} \\
{\tt eivanov@theor.jinr.ru}\\

\end{center}
\vspace{0.2cm}
\vskip 0.6truecm  \nopagebreak

\begin{abstract}
\noindent This talk summarizes the study of superfield gaugings of isometries
of extended supersymmetric mechanics in {\tt hep-th/0605211},
{\tt hep-th/0611247} and {\tt arXiv:0706.0706}.
The gauging procedure provides a manifestly supersymmetric realization of
$d=1$ automorphic dualities which interrelate various irreducible off-shell multiplets
of $d=1$ extended supersymmetry featuring the same
number of physical fermions but different divisions of bosonic fields into the physical and auxiliary
subsets. We concentrate on the most interesting ${\cal N}=4$ case and demonstrate
that, with a suitable choice of the symmetry to be gauged, all such multiplets of ${\cal N}=4$ supersymmetric
mechanics and their generic superfield actions can be obtained  from the ``root'' multiplet ${\bf (4,4,0)}$
and the appropriate gauged subclasses of the generic superfield action of the latter by
a simple universal recipe.
\end{abstract}
\vspace{4cm}

\begin{center}
\noindent{\it Talk at the International Workshop SQS'07,
July 30 - August 4, 2007, Dubna, Russia}
\end{center}
\newpage

\end{titlepage}

\section{Introduction}

In recent years, there has been a continuous effort to improve our understanding
of $d=1$ supersymmetric models (supersymmetric mechanics).  These studies may be motivated as follows:
\begin{itemize}
\item{The $d=1$ models form a testing ground for higher-dimensional theories.}
\item{The search for supersymmetric extensions of some well known $d=1$ models such as the Calogero
model requires a good knowledge of the basic structures of extended supersymmetry in $d=1$.}
\item{Among all supersymmetric models, of particular interest are the superconformal models,
which play a role in black hole physics and AdS$_2$/CFT$_1$ version of the general AdS/CFT correspondence.}
\end{itemize}
We would like to emphasize that not all properties of $d=1$ supersymmetric models can be recovered by reduction
from higher dimensions. Surprisingly, there come out some new features which are specific just for the $d=1$ case.
First of all, this concerns the structure of the irreducible off-shell multiplets of extended supersymmetry in $d=1$.
These multiplets are characterized by three integer numbers $({\bf n}_{1}, {\bf n}_{2}, {\bf n}_{3})$ where ${\bf n}_{1}$
is the number of physical bosons (a boson will be considered as physical
if it appears with a second-order time derivative in the field equations),
${\bf n}_{2}$ is the number of physical fermions (entering the field equations with a first-order time derivative)
and ${\bf n}_{3} = {\bf n}_{2} -{\bf n}_{1}$ is that of auxiliary bosons (possessing algebraic
field equations). In most physically interesting cases ${\bf n}_{2}$ coincides with the number ${\cal N}$ of supercharges.
Not all of these off-shell multiplets can be obtained as a reduction of those of the $d>1$
supersymmetries\footnote{An obvious exception is the $d=2$ heterotic supersymmetry, the multiplet structure
of which is basically the same as in one dimension.}. Some of them have as the $d>1$ counterparts essentially
on-shell multiplets (modulo a possibility of adding an infinite number of auxiliary fields as in the
harmonic superspace approach). Many linear $d=1$ multiplets have nonlinear ``cousins'' on which the relevant
$d=1$ supersymmetry is realized in an intrinsically nonlinear fashion.

In this report we shall focus on the case of ${\cal N}=4$ supersymmetric
mechanics\footnote{The multiplets and actions of the supersymmetric
mechanics with $N<4$ can be obtained as proper reductions or truncations of the generic ${\cal N}=4$ models.}.
The list of ${\cal N}=4, d=1$  multiplets ranges
from the multiplet ${\bf (4,4,0)}$ with the maximal number ${\bf 4}$ of physical bosons to the fermionic
multiplet ${\bf (0,4,4)}$ having no physical bosons at all. These multiplets were described in \cite{GR0,GR1,GR}
and also in \cite{PT}. It was observed that they are related to each other by the  so-called $d=1$
{\it automorphic dualities}
\be
{\bf (4,4,0)} \leftrightarrow {\bf (3,4,1)} \leftrightarrow {\bf (2,4,2)} \leftrightarrow  \cdots\,. \lb{I}
\ee
These relationships were established in \cite{GR0}-\cite{PT} at the linear level of free actions.
Further work on these multiplets and the relationships between them was carried out in \cite{IL,IKL1,root,To}.
Of particular relevance is the article \cite{root} where, at the component level,
many results related to the linear and nonlinear automorphic dualities were summarized and the role
of the ``root'' multiplet ${\bf (4,4,0)}$ as the generating one for other ${\cal N}=4$ multiplets was pointed out.

Recently, in a series of three articles \cite{I,DI2,DI3}, we described a systematic superfield way
to relate various ${\cal N}=4$ multiplets using the procedure of gauging isometries in superspace. The present report
is a brief summary of this superfield gauging procedure. Before turning to the subject,
let us recall a few facts about ${\cal N}=4, d=1$ superspace.

\section{${\cal N}=4$ Superspace}
\subsection{Standard superspace}
${\cal N}=4, d=1$ superspace is parametrized by the time $t$ and four Grassmann variables
which may be organized in two complex doublets $\theta^{i}, \bar\theta_{i}$, $(i=1,2)$ with respect to
one of the two commuting $SU(2)$ automorphism groups of the ${\cal N}=4, d=1$ Poincar\'e superalgebra. The
covariant spinor derivatives are given by
\begin{equation}
D_{i}=\frac{\partial}{\partial\theta^{i}}+i\bar\theta_{i}\partial_{t}\,,
\quad \bar D^{i}=\frac{\partial}{\partial\bar\theta_{i}}+i\theta^{i}\partial_{t}\,,
\quad \{D_{i},\bar D^{j}\}=2i\delta^j_{i}\partial_{t}\,.
\end{equation}
Various ${\cal N}=4$ multiplets may be described in superspace using the following superfields.
\begin{itemize}
\item{${\bf (4,4,0)}\,$: $\quad q^{ia}(t,\theta,\bar\theta)\,,\quad i=1,2,\quad a=1,2$\\
In order to contain the right number of fermionic fields, this superfield should satisfy
the supersymmetric constraints:
\begin{equation}
D^{(i}q^{j)a}=0\,,\quad \bar D^{(i}q^{j)a}=0\,.\label{troc}
\end{equation}
The component fields contained in this superfield are four bosons
$f^{ia}(t)=q^{ia}\vert_{\theta=\bar\theta=0}$ and four fermions
$\chi^{a}= 1/2\, D_{i}q^{ia}\vert_{\theta=\bar\theta=0}\,$, $\bar\chi^{a}=1/2 \,\bar D_{i}q^{ia}\vert_{\theta=\bar\theta=0}\,$.}
\item{${\bf (3,4,1)}\,$: $\quad W^{(ij)}(t,\theta,\bar\theta)$\\
Constraints:
\begin{equation}
D^{(i}W^{jk)}=0\,,\quad \bar D^{(i}W^{jk)}=0\,.\label{troc1}
\end{equation}}
\item{${\bf (2,4,2)}\,$: $\quad \Phi(t,\theta,\bar\theta), \quad \bar\Phi(t,\theta,\bar\theta)$\\
Constraints: $\bar D^{i}\Phi=0$ ($\Phi$ is a chiral superfield)}
\item{${\bf (1,4,3)}\,$: $\quad \Omega(t,\theta,\bar\theta)$\\
This real superfield already features the right number of physical bosons and fermions.
The right number of auxiliary fields is guaranteed by the second
order constraints:
\begin{equation}
D^{i}D_{i}\,\Omega=0\,,\quad \bar D^{i}\bar D_{i}\,\Omega=0\,,
\quad [D^{i},\bar D_{i}]\,\Omega=0\,.\label{consc}
\end{equation}}
\item{${\bf (0,4,4)}\,$: $\quad \psi^{ia}(t,\theta,\bar\theta)$ is fermionic and satisfies
the same constraints as in (\ref{troc}).}
\end{itemize}

One can also define ``mirror'' counterparts of these multiplets, with another
$SU(2)$ factor of the full ${\cal N}=4, d=1$ automorphism group $SO(4) \sim SU(2)\times SU(2)$ being manifest.
These two sets of irreducible off-shell ${\cal N}=4, d=1$ multiplets are actually distinguishable only if they are considered pairwise.
For simplicity,  we shall consider only the above set.

\subsection{Harmonic superspace}
A convenient way to solve constraints of the type (\ref{troc}) is to use the harmonic superspace
approach \cite{HSS,HSS1}. It consists in adding
new {\it harmonic} coordinates $u^{\pm}_{i}$ such that
\begin{equation}
\left(\begin{array}{cc} u^{+}_{1} &u^{-}_{1}\\u^{+}_{2} &u^{-}_{2}\end{array}\right)
\in SU(2)\quad\Rightarrow u^{+i}u^{-}_{i}=1\,,\quad u^{+i}=\overline{u^{-}_{i}}\,.
\end{equation}
Since new coordinates have been introduced, one can also define new derivatives
\begin{eqnarray}
& D^{++}=u^{+}_{i}\frac{\partial}{\partial u^{-}_{i}}\,,
\quad D^{--}=u^{-}_{i}\frac{\partial}{\partial u^{+}_{i}}\,,\quad D^{0}
=u^{+}_{i}\frac{\partial}{\partial u^{+}_{i}}-u^{-}_{i}\frac{\partial}{\partial u^{-}_{i}}\,,&\nonumber\\&
[D^{0},D^{\pm\pm}]=\pm 2D^{\pm\pm}\,,\quad [D^{++},D^{--}]=2D^{0}\,.&
\end{eqnarray}
Let us project the superfield $q^{ia}$ and the derivatives $D^{i}$, $\bar D^{i}$ on the doublet $u^{+}_{i}$
\begin{equation}
q^{+a}=q^{ia}u^{+}_{i}\,,\quad D^{+}=D^{i}u^{+}_{i}\,,
\quad \bar D^{+}=\bar D^{i}u^{+}_{i}\,.\label{analyt}
\end{equation}
The constraints (\ref{troc}) now can be equivalently rewritten as
\be
D^{+}q^{+a}=0\,,\quad \bar D^{+}q^{+a}=0\,. \lb{qAnC}
\ee
Superfields satisfying these constraints are called analytic. As in the case of chiral superfields,
the meaning of these constraints is that the superfield $q^{+a}$ depends only on half
of the odd coordinates of harmonic superspace, $q^{+a} = q^{+ a}(t_A, \theta^+, \bar\theta^+, u^\pm)\,, \,
\theta^+ =\theta^iu^+_i,\bar\theta^+ = \bar\theta^iu^+_i\,$. The superfield $q^{+a}$ also depends
on the new harmonic coordinates $u^{\pm}_{i}$. However, according to (\ref{analyt}),
this dependence is very restricted, being just linear. This restriction can be concisely reformulated
as the harmonic constraint:
\begin{equation}
D^{++}q^{+a}=0\,. \lb{qHarC}
\end{equation}
The analyticity conditions \p{qAnC} supplemented with the harmonic constraint \p{qHarC} yield an equivalent
superfield formulation of the multiplet ${\bf (4,4,0)}$.

The multiplet ${\bf (3,4,1)}$ also has a simple description in harmonic superspace. Again,
we project the superfield $W^{ij}$ on $u^{+}_i\,$:
\begin{equation}
W^{++}=W^{ij}u^{+}_{i}u^{+}_{j}\,.\label{def}
\end{equation}
Then the constraints (\ref{troc1}) and the definition (\ref{def}) amount in harmonic superspace to
$$ D^{+}W^{++}=0\,,\quad \bar D^{+}W^{++}=0 \; \Rightarrow \; W^{++}= W^{++}(t_A, \theta^+, \bar\theta^+, u^\pm)\quad(\mbox{{\sl analyticity constraint}})\,,$$
$$D^{++}W^{++}=0 \quad(\mbox{{\sl harmonic constraint}})\,.$$

Two remarks are in order :
\begin{itemize}
\item{All constraints written so far are off-shell, i.e. they do not restrict
the time dependence of the surviving fields. In $d=4$, the same kind of constraints would lead to field equations.}
\item{All constraints written so far are linear in superfields. However, as was already mentioned, there
exist nonlinear extensions of these constraints. This is a peculiarity of $d=1$.}
\end{itemize}
\subsection{Free actions}
Let us give the superspace and component forms of the free action of the above multiplets.
\begin{itemize}
\item{${\bf (4,4,0)}\,$:  $\quad $
$S_{0}=\int dtd^{4}\theta du\, q^{+a}D^{--}q^{+}_{a}
\sim \int dt\,(\dot f^{ai}\dot f_{ai}+i\bar\chi^{a}\dot\chi_{a})$\,.}
\item{${\bf (3,4,1)}\,$: $\quad S_{0}=\int dtd^{4}\theta du\, W^{++}(D^{--})^{2}W^{++} \sim \int dt\,(\dot w^{(ij)}\dot w_{(ij)}
+i\bar\lambda^{i}\dot\lambda_{i}+f^{2})$\,.}
\item{${\bf (2,4,2)}\,$: $\quad S_{0}=\int dtd^{4}\theta\,\bar\Phi\Phi
\sim \int dt\,(\dot{\bar\phi}\dot\phi+i\bar\tau^{i}\dot\tau_{i}+\bar F F)$\,.}
\item{${\bf (1,4,3)}\,$:  $\quad S_{0}=\int dt d^4{\theta}\,\Omega^2
\sim \int dt\,(\dot\omega\dot\omega+i\bar\rho^{i}\dot\rho_{i}+f^{(ij)}f_{(ij)})$\,.}
\end{itemize}
Here the symbol $\sim$ means ``up to a numerical renormalization constant''.
\section{Gauging a symmetry}
Now we are prepared to show that the chain of the automorphic dualities \p{I} amounts to gauging certain
symmetries implementable on the multiplet ${\bf (4,4,0)}$ and choosing the appropriate gauge-fixing conditions.
\subsection{Transformations of the ``root'' (4,4,0) multiplet}
As stated above, the ${\bf (4,4,0)}$  multiplet is described
in harmonic superspace by a superfield $q^{+a}$, $a=1,2,$ satisfying the constraints \p{qAnC} and \p{qHarC}.
We shall now need transformations of this superfield which are symmetries of the classical action.
As a first step, we require that these transformations leave invariant the constraints.
This is the case for the following transformations
\begin{enumerate}
\item{
{\sl Shift}: $\quad \delta_{1}q^{+a}=\lambda_{1}{m^{a}}_{b}u^{+b}\,$.}
\item{ $SU(2)$ {\sl rotations}: $\quad \delta_{su(2)}q^{+a}=\lambda^{a}_{\;\;b}\,q^{+b}\,,
\;\lambda^{a}_{a} = 0\,$.}
\item{ $U(1)\subset SU(2)$ {\sl rotation}: $\quad \delta_{2}q^{+a}=\lambda_{2}{c^{a}}_{b}q^{+b}\,,\quad
{c^{a}}_{a}=0\,$.}
\item{{\sl Scale transformation}: $\quad \delta_{3}q^{+a}=\lambda_{3}q^{+a}\,$.}
\end{enumerate}
The transformations 1, 2 and 3 are invariances of the free action of the ${\bf (4,4,0)}$ multiplet.
Requiring invariance under the rescalings 4 picks up a more complicated action, with a non-trivial bosonic target space
metric.
\subsection{An example: gauging a shift symmetry}
To explain the basic idea, we specialize to the case of a shift symmetry with ${m^{a}}_{b}={\delta^{a}}_{b}$.
The gauging procedure as its first step involves replacing the global parameter $\lambda_{1}$
by a superfield $\Lambda_{1}(t,\theta,\bar\theta,u)$ which depends on the coordinates of harmonic superspace.
We require the local transformations to respect the analyticity, and thus $\Lambda_{1}$
is an analytic superfield
\begin{equation}
\delta_{1}q^{+a}=\Lambda_{1}u^{+a}\,,\qquad D^{+}\Lambda_{1}=0\,,\quad \bar D^{+}\Lambda_{1}=0\; \Leftrightarrow\;
\Lambda_{1} = \Lambda_{1}(t_A, \theta^+, \bar\theta^+, u^\pm)\,.\label{gauge}
\end{equation}
The harmonic constraint needs to be covariantized. This can be done by introducing
an analytic gauge superfield $V^{++}(t_A, \theta^+, \bar\theta^+, u^\pm)$ with the gauge transformation law
$$\delta_{1}V^{++}=D^{++}\Lambda_{1}\,.$$
Then, the covariantized harmonic constraint reads
$$\nabla^{++}q^{+a}=D^{++}q^{+a}-V^{++}u^{+a}=0\,.$$
The $D^{--}$ derivative also needs to be covariantized. We introduce a non-analytic superfield $V^{--}$
with the gauge transformation $\delta_{1}V^{--}=D^{--}\Lambda_{1}\,$. The covariant derivative of $q^{+a}$ reads
$$\nabla^{--}q^{+a}=D^{--}q^{+a}-V^{--}u^{+a}\,.$$
Since the superfield parameter $\Lambda_{1}$ has zero charge, the derivative $D^{0}$
needs not be covariantized. We then have the algebra
$$[\nabla^{++},\nabla^{--}]=D^{0}\;\Rightarrow\; D^{++}V^{--}-D^{--}V^{++}=0\,.$$
This equation determines $V^{--}$ in terms of $V^{++}\,$.
The covariantization of the free action is
\begin{equation}
S_{g}=\int dtd^{4}\theta du\,q^{+a}\nabla^{--}q^{+}_{a}\,.
\end{equation}
The gauge transformation (\ref{gauge}) implies
$$\delta_{1}\left(q^{+a}u^{-}_{a}\right)=\Lambda_{1}\,.$$
Thus we may choose a supersymmetric unitary gauge such that
\begin{equation}
q^{+a}u^{-}_{a}=0\,. \lb{Gauge}
\end{equation}
Then, what remains from the superfield $q^{+a}$ is the projection $W^{++}=q^{+a}u^{+}_{a}\,$.
The harmonic constraint expresses $V^{++}$ in terms of $W^{++}\,$,
and also properly constrains $W^{++}$
\begin{equation}
\nabla^{++}q^{+a}=0\,\Rightarrow\,\,
 \left\{\begin{array}{c}V^{++}=W^{++}\\D^{++}W^{++}=0\end{array}\right. .
\end{equation}
We recognize $W^{++}$ as the superfield providing the harmonic superspace description of the ${\bf (3,4,1)}$ multiplet.
The gauge invariant action $S_{g}$ reduces to the free ${\bf (3,4,1)}$ action.

Instead of a supersymmetric gauge, we might equally choose the Wess-Zumino (WZ) gauge
\begin{equation} V^{++}= 2i\theta^{+}\bar\theta^{+}A(t)\,,\end{equation}
the only surviving component in $V^{++}$ being the gauge field $A(t)$.
Notice the very unusual fact that only {\it one} bosonic field $A(t)$ remains in the supermultiplet in the WZ gauge.
The reason why this is possible and compatible with supersymmetry is explained in \cite{I}.
The residual gauge freedom of the component fields is given by
\begin{equation}
\delta_{1}A(t)= -\partial_{t}\lambda_{1}(t)\,,\quad \delta_{1}f^{ia}(t)=\lambda_{1}\epsilon^{ai}\,,
\quad \delta_{1}\chi^{a}(t)=0\,, \quad \lambda_1 = \Lambda_1\vert_{\theta =\bar\theta =0}\,. \lb{Resid}
\end{equation}
In WZ gauge, the gauge invariant action $S_{g}$ becomes, in terms of components,
\begin{equation}
S_{g} \sim \int dt\left[(\dot f^{ia}-A\epsilon^{ia})(\dot f_{ia} +A\epsilon_{ia})+i\bar\chi^{a}\dot\chi_{a}\right].
\end{equation}
The essential degrees of freedom are revealed by imposing the further (unitary) gauge
\begin{equation}
\delta_{1}(f^{ia}\epsilon_{ia})= 2\lambda_{1}(t)\,,\,\, \Rightarrow \;
\mbox{{\sl unitary gauge} :}\,\, f^{ia}\epsilon_{ia}=0\,.
\end{equation}
The action then becomes
\begin{equation}
S_{g} \sim \int dt\left[\dot f^{(ia)}\dot f_{(ia)}+i\bar\chi^{a}\dot\chi_{a}+ 2 A^2\right].
\end{equation}
The remaining fields are a triplet of physical bosons $f^{(ia)}$, a complex doublet
of fermions $\chi^{a}$, $\bar\chi_{a}$ and an auxiliary field $A$. This is just the component
content of the ${\bf (3,4,1)}$ supermultiplet. A Higgs-type phenomenon has occurred, the gauge field
has ``eaten'' a Goldstone boson to become an auxiliary field.

In order to reproduce the most general sigma-model type superfield action of the multiplet ${\bf (3,4,1)} \leftrightarrow  W^{++}$,
one should start from the general superfield $q^+$ action invariant under the shifts \p{gauge} and
pass to the gauged action by the same rules as above.
\section{General results}
Here we sketch the results of applying the gauging procedure to other $q^{+a}$ symmetries listed in Sect.3.1.
Details can be found in \cite{I} - \cite{DI3}.
\subsection{Cases considered}
As follows from the simple example above, the number of physical bosons which disappear in the process of gauging is equal
to the number of gauge symmetries. We have studied the cases
\begin{itemize}
\item{{\sl One isometry} \cite{I}: $\quad {\bf (4,4,0)}\,\Rightarrow\,{\bf (3,4,1)}\,$.}
\item{{\sl Two isometries} \cite{DI3}: $\quad {\bf (4,4,0)}\,\Rightarrow\,{\bf (2,4,2)}\,$.}
\item{{\sl Three isometries} \cite{I,DI2}: $\quad {\bf (4,4,0)}\,\Rightarrow\,{\bf (1,4,3)}\,$.}
\end{itemize}
We have considered the case of general interacting $q^+$ Lagrangians, which may be interpreted as describing
the motion of a point particle on a curved manifold. Moreover, the point particle may be
interacting with an external magnetic field. If an abelian symmetry is gauged,
it also becomes possible to generate a potential term in the final action from a Fayet-Iliopoulos term
$$\sim \,\int dt_{A}d\theta^{+}d\bar\theta^{+}du\, V^{++}$$
combined with the superfield $q^{+a}$ coupling to an external magnetic field.
\subsection{One isometry \cite{I}: ${\bf (4,4,0)}\,\Rightarrow\,{\bf (3,4,1)}$}
We distinguish the options related to three different one-generator groups listed in Sect.3.1
\begin{itemize}
\item{{\sl Shift isometry}:\\
One can obtain the general action of the linear ${\bf (3,4,1)}$ multiplet (including both the sigma-model type and
superpotential type terms) in two different ways, starting from
\begin{itemize}
\item{Linear harmonic constraints \p{qHarC} and a general shift-invariant action of the corresponding analytic
${\bf (4,4,0)}$ superfield $q^{+a}\,$;}
\item{General non-linear shift-invariant harmonic constraints generalizing \p{qHarC} and a sum of the ``free'' superfield bilinear
$q^{+ a}$ action\footnote{This action has the same form as that given in Sect.2.3 but it yields a non-trivial sigma-model type
interaction after passing to components due to the non-linearity of the underlying superfield harmonic constraint. In particular,
the bosonic sector of this action is the $d=1$ pullback of the general 4-dimensional hyper K\"ahler squared interval with one triholomorphic
isometry (the Gibbons-Hawking ansatz).} and a general shift-invariant coupling of $q^{+ a}$ to an external magnetic field.}
\end{itemize}}
\item{{\sl Rotational isometry}:\\
One starts from the subclass of general $q^{+ a}$ actions which enjoys invariance under the transformations 3 defined in Sect.3.1
(this particular action still contains both the sigma-model and superpotential parts)
and gauges this isometry by the analytic superfield $V^{++}$, quite analogously to the shifting case worked out in Sect.3.2. The
general $W^{++}$ action is reproduced in the supersymmetric unitary gauge analogous to \p{gauge} under the identification
$W^{++} \sim q^{+ a}c_{ab}q^{+b}\,$. As opposed to the case of a shifting isometry, the genuine free ${\bf (4,4,0)}$
action already leads to an interacting ${\bf (3,4,1)}$ action. Moreover, the FI term directly
yields the known \cite{IKL0,IL}
conformally invariant ${\bf (3,4,1)}$ potential term.}
\item{{\sl Scale isometry}:\\
Once again, one starts from the appropriate invariant (under the transformations 4 of Sect.3.1) subclass of the generic $q^{+ a}$ actions.
The new feature of this case is that in the unitary-type gauge the linear constraint \p{qHarC}
leads to the non-linear ${\bf (3,4,1)}$ constraint
$$D^{++}W^{++}+W^{++}W^{++}=0\,.$$ The outcome is the most general superfield action of this nonlinear
${\bf (3,4,1)}$ multiplet\footnote{More general types of nonlinear ${\bf (3,4,1)}$ multiplets can be obtained, starting from
$q^{+a}$ subjected to some nonlinear harmonic constraints which are still covariant under the scale transformations \cite{DIprep}.}.}
\end{itemize}
\subsection{Two isometries \cite{DI3}: ${\bf (4,4,0)}\,\Rightarrow\,{\bf (2,4,2)}$}
The various cases which have been studied are
\begin{itemize}
\item{{\sl Two shift isometries (abelian)}:\\
The linear ${\bf (4,4,0)}$ multiplet leads to a linear twisted chiral ${\bf (2,4,2)}$ multiplet\footnote{It becomes
the standard chiral ${\cal N}=4, d=1$ superfield after switching to the alternative basis in ${\cal N}=4, d=1$ superspace,
with another automorphism $SU(2)$
group $\subset SO(4) \sim SU(2)\times SU(2)\,$ being manifest.}
\be D^1\phi=0\,,\quad \bar D^1\phi=0\,. \lb{Ch}
\ee
}
\item{{\sl  One rotational and one scale isometries (abelian)}:\\
The linear ${\bf (4,4,0)}$ multiplet leads to a non-linear twisted chiral ${\bf (2,4,2)}$ multiplet
\be D^1\phi+\phi D^2\phi=0\,,\quad \bar D^1\phi+\phi\bar D^2\phi=0\,. \lb{NonCh}
\ee
}
\item{{\sl One shift and one scale isometries (non-abelian)}:\\
Again, the linear ${\bf (4,4,0)}$ multiplet leads to a non-linear twisted chiral ${\bf (2,4,2)}$ multiplet.
Notice that the relevant Killing vectors
${\cal T}_{1}=u^{+a}\frac{\partial}{\partial q^{+a}}$ (the shift isometry) and ${\cal T}_{3}
=q^{+a}\frac{\partial}{\partial q^{+a}}$ (the scale isometry) form a non-abelian solvable algebra
$$\left[{\cal T}_{1},{\cal T}_{3}\right]={\cal T}_{1}\,.$$}
\end{itemize}

Note that in the case of two isometries it turns out advantageous to use the bridge between
the analytic gauge group (the so called $\lambda$ group) and the gauge group with a parameter independent of harmonics
(the so called $\tau$ group). Let us describe
this bridge in the case of a scale isometry. The gauge-covariantized harmonic constraint on the ``root'' multiplet reads
$$(D^{++}-V^{++})q^{+a}=0\,.$$
We introduce the non-analytic gauge superfield $v$ such that
$$V^{++}=D^{++}v\,,\quad \delta v=\Lambda_{3}(t,\theta,\bar\theta,u)
+\tau_{3}(t,\theta,\bar\theta)\,.$$
Then the non-analytic superfield $Q^{+a}=e^{-v}q^{+a}$ satisfies a simple harmonic condition
\be D^{++}Q^{+a}=0\,\,\Rightarrow\,\, Q^{+a}(t,\theta,\bar\theta,u)
=Q^{ia}(t,\theta,\bar\theta)u^{+}_{i}\,, \; \delta Q^{ia} = -\tau_3 Q^{ia}\,. \lb{Br}
\ee
The price to pay is that the Grassmann analyticity constraint on $q^{+a}$, being rewritten in terms of  $Q^{+a}$,
displays a connection term. Four ordinary ${\cal N}=4, d=1$ superfields $Q^{ia}$ satisfying the corresponding nonlinear
version of the analyticity constraint and undergoing $\tau_3$ gauge transformation, eqs. \p{Br}, provide an alternative description
of the nonlinear ${\bf (3,4,1)}$ multiplet in the ordinary ${\cal N}=4, d=1$ superspace. Gauging away one of these superfields using
the $\tau_3$ gauge freedom leaves us with 3 superfields satisfying some nonlinear constraint. In the case of two isometries
one has two independent $\tau$ gauge parameters which are capable to gauge away two out of four superfields. The remaining
two superfields satisfy the linear or nonlinear chirality constraints \p{Ch} or \p{NonCh}.
\subsection{Three isometries \cite{I,DI2}: ${\bf (4,4,0)}\,\Rightarrow\,{\bf (1,4,3)}$}
We have considered the cases of three mutually commuting shift transformations (abelian symmetry group, item 1 in Sect.3.1,
with $m^a_a =0$) and of three rotations (non-abelian symmetry group $SU(2)$, the item 2 in Sect.3.1).
In both cases, the general gauge invariant action
of the linear ${\bf (4,4,0)}$ multiplet leads to the general action of the multiplet ${\bf (1,4,3)}\,$.
A peculiar feature  of this construction is that the ${\bf (1,4,3)}$ superfield $\Omega$
satisfying the constraints (\ref{consc}) is obtained from an analytic gauge prepotential $V$ through the formula
\be \Omega=\int du\,V\,, \quad \delta V=D^{++}\Lambda^{--}\,, \lb{OV}
\ee
where $\Lambda^{--}$ is an analytic superfield gauge parameter. By definition, $\Omega$ is gauge invariant.

Only the non-abelian gauging allows one to keep track of the superconformal properties of all involved superfields.
In this case, the gauging procedure preserves the superconformal $D(2,1;\alpha)$ $(\alpha \neq 0)$ covariance
of the harmonic ${\bf (4,4,0)}$ constraint. This property is conducive to the existence of a new mechanism to
generate a conformal potential term for the ${\bf (1,4,3)}$ multiplet via a superconformal coupling of the latter
to the fermionic multiplet ${\bf (0,4,4)}\,$.
\section{Conclusion}
In summary, all known ${\cal N}=4, d=1$ off-shell multiplets in the superfield description can be obtained from the ``root''
analytic ${\bf (4,4,0)}$ superfield $q^{+ a}$ by gauging some symmetries realized on $q^{+ a}$. This provides
a manifestly supersymmetric formulation of the ``$d=1$ automorphic duality''.  Linear as well
as non-linear multiplets can be obtained in this way, depending on the choice of the symmetry group
to be gauged. Among the directions of further study it is worth mentioning an extension of the gauging procedure to the case of
superfields $q^{+a}$ subjected to the most general nonlinear harmonic constraints \cite{DIprep}, an analysis of implications
of this procedure for the superfield actions involving both the standard and ``mirror'' ${\cal N}=4$ multiplets and, finally, a generalization
to the case of ${\cal N}=8$ supersymmetric mechanics, with the ``root'' multiplet ${\bf (8,8,0)}\,$ \cite{N8,To}. On this way, we hope to discover new
models of supersymmetric mechanics and to get deeper insights into the geometric and algebraic structure of the known models.

\subsection*{Acknowledgments}
E.I. acknowledges a partial support from RFBR grants, projects No
03-02-17440 and No 04-02-04002, the grant
INTAS-00-00254, the DFG grant No.436 RUS 113/669-02, and a grant of
the Heisenberg-Landau program.


\begin{thebibliography}{99}
\bibitem{GR0} S.J.~Gates, Jr., L.~Rana, ``On Extended Supersymmetric
Quantum Mechanics'', Maryland Univ. Preprint \# UMDPP 93-24, Oct. 1994.
\bibitem{GR1} S.J.~Gates, Jr., L.~Rana, Phys. Lett. B345 (1995) 233,
{\tt hep-th/9411091}.
\bibitem{GR}S.J.~Gates, Jr., L.~Rana, Phys. Lett. B342 (1995) 132,
{\tt hep-th/9410150}.
\bibitem{PT} A.~Pashnev, F.~Toppan, J. Math. Phys. 42 (2001) 5257,
{\tt hep-th/0010135}.
\bibitem{IL} E.~Ivanov, O.~Lechtenfeld, JHEP 0309 (2003) 073,
{\tt hep-th/0307111}.
\bibitem{IKL1}E.~Ivanov, S.~Krivonos, O.~Lechtenfeld,
Class. Quant. Grav. 21 (2004) 1031, {\tt hep-th/0310299}.
\bibitem{root} S.~Bellucci, S.~Krivonos, A.~Marrani, E.~Orazi,
Phys. Rev. D73 (2006) 025011, {\tt hep-th/0511249}.
\bibitem{To} Z.~Kuznetsova, M.~Rojas, F.~Toppan, JHEP 0603 (2006) 098, {\tt  hep-th/0511274}.
\bibitem{I} F.~Delduc, E.~Ivanov, Nucl. Phys. B753 (2006) 211, {\tt hep-th/0605211}.
\bibitem{DI2} F.~Delduc, E.~Ivanov, Nucl. Phys. B770 (2007) 179, {\tt hep-th/0611247}.
\bibitem{DI3}F.~Delduc, E.~Ivanov, Nucl. Phys. B787 (2007) 176, {\tt arXiv:0706.0706[hep-th]}.
\bibitem{HSS}A.~Galperin, E.~Ivanov, V.~Ogievetsky, E.~Sokatchev,
Pis'ma ZhETF 40 (1984) 155 [JETP Lett. 40 (1984) 912];
A.S.~Galperin, E.A.~Ivanov, S.~Kalitzin, V.I.~Ogievetsky, E.S.~Sokatchev,
Class. Quant. Grav. 1 (1984) 469.
\bibitem{HSS1} A.S.~Galperin, E.A.~Ivanov, V.I.~Ogievetsky, E.S.~Sokatchev,
``{\it Harmonic Superspace}'', Cambridge University Press 2001, 306 p.
\bibitem{IKL0} E.~Ivanov, S.~Krivonos, O.~Lechtenfeld, JHEP 0303 (2003) 014,
{\tt hep-th/0212303}.
\bibitem{DIprep} F.~Delduc, E.~Ivanov, work in preparation.
\bibitem{N8}S.~Bellucci, E.~Ivanov, S.~Krivonos, O.~Lechtenfeld,
Nucl. Phys. B699 (2004) 226, {\tt hep-th/0406015}; E.~Ivanov, O.~Lechtenfeld, A.~Sutulin,
Nucl. Phys. B790 (2008) 493, {\tt arXiv:0705.3064[hep-th]}.

\end{thebibliography}
\end{document}